%% file: Main.tex
\documentclass[journal]{IEEEtran}

\usepackage{xcolor}
\usepackage{caption}
\usepackage{relsize}
\usepackage{booktabs}
\usepackage{amsmath}
\usepackage{amsfonts}

\ifCLASSINFOpdf
   \usepackage[pdftex]{graphicx}
  \graphicspath{{../figures/}}
   \DeclareGraphicsExtensions{.pdf,.jpeg,.png}
\else
   \usepackage[dvips]{graphicx}
   \graphicspath{{../figures/}}
   \DeclareGraphicsExtensions{.eps}
\fi
\ifCLASSOPTIONcompsoc
  \usepackage[caption=false,font=normalsize,labelfont=sf,textfont=sf]{subfig}
\else
  \usepackage[caption=false,font=footnotesize]{subfig}
\fi
\usepackage{url}

\usepackage{xcolor}

\usepackage{tikz}

\newcommand\copyrighttext{%
  \footnotesize This work has been submitted to the IEEE for possible publication. Copyright may be transferred without notice, after which this version may no longer be accessible}
\newcommand\copyrightnotice{%
\begin{tikzpicture}[remember picture,overlay]
\node[anchor=south,yshift=10pt] at (current page.south) {\fbox{\parbox{\dimexpr\textwidth-\fboxsep-\fboxrule\relax}{\copyrighttext}}};
\end{tikzpicture}%
}

\begin{document}

%
\title{Non-Contrastive Learning-based Behavioural Biometrics for Smart IoT Devices}

\author{
		Oshan Jayawardana,
		Fariza Rashid, and
		Suranga Seneviratne
\thanks{ O. Jayawardana is with The School of Computer Science, The University Sydney, Australia and The University of Moratuwa, Sri Lanka.}
\thanks{ F. Rashid and S. Seneviratne are with The School of Computer Science, The University Sydney, Australia.}}

\maketitle

\begin{abstract}
Behaviour biometrics are being explored as a viable alternative to overcome the limitations of traditional authentication methods such as passwords and static biometrics. Also, they are being considered as a viable authentication method for IoT devices such as smart headsets with AR/VR capabilities, wearables, and erables, that do not have a large form factor or the ability to seamlessly interact with the user. Recent behavioural biometric solutions use deep learning models that require large amounts of annotated training data. Collecting such volumes of behaviour biometrics data raises privacy and usability concerns. To this end, we propose using SimSiam-based non-contrastive self-supervised learning to improve the label efficiency of behavioural biometric systems. The key idea is to use large volumes of unlabelled (and anonymised) data to build good feature extractors that can be subsequently used in supervised settings. Using two EEG datasets, we show that at lower amounts of labelled data, non-contrastive learning performs 4\%--11\% more than conventional methods such as supervised learning and data augmentation. We also show that, in general, self-supervised learning methods perform better than other baselines. Finally, through careful experimentation, we show various modifications that can be incorporated into the non-contrastive learning process to archive high performance. 
\end{abstract}

\copyrightnotice

\input sections/Abstract
\input sections/Introduction
\input sections/UpdatedRelated
\input sections/Methodology

\input sections/Results
\input sections/ResultsAnalysis
\input sections/Conclusion
\vspace{-3mm}
\bibliographystyle{IEEEtran}
\bibliography{IEEEabrv,biblio.bib}
\end{document}

%% file: sections/Abstract.tex

%% file: sections/Introduction.tex
\section{Introduction}
\label{sec:introduction}









The pervasive use of smart devices and the vast amounts of sensitive information stored in those devices exacerbate the problem of user authentication on smart devices. Traditional methods such as passwords, PINs, and security tokens have usability issues~\cite{weaver2006biometric} and static biometrics such as fingerprinting and face ID are vulnerable to spoofing attacks. As a result, \textit{behavioural biometrics} has been explored by many works as a user-friendlier (i.e., implicit by nature and no extra effort is required from the user) and secure (i.e., difficult to spoof and allows continuous authentication) alternative for user authentication in smart devices. Example behavioural biometric modalities include gait patterns~\cite{5638036}), typing patterns~\cite{banerjee2012biometric,killourhy2009comparing}, breathing acoustics~\cite{chauhan2017breathprint,chauhan2018performance}, and EEG patterns~\cite{sooriyaarachchi2020musicid,zhang2018mindid}. Behavioural biometrics also finds applications in Smart IoT devices that either do not have enough form factor or limited interactive components~\cite{sooriyaarachchi2020musicid,chauhan2018performance}


{Compared to static biometrics, behaviour biometrics needs a significant number of training samples to be collected from users at registration time and in most cases at different contextual settings~\cite{chauhan2017breathprint, sooriyaarachchi2020musicid}. Moreover, the majority of recent behavioural biometrics solutions use deep learning models that are known to require higher amounts of training data~\cite{abuhamad2020sensor,miller2021using}. Specifically, many solutions used Convolutional Neural Networks (CNNs)~\cite{gadaleta2018idnet, chauhan2018performance} or Recurrent Neural Networks (RNNs)~\cite{zhang2018mindid, chauhan2018breathing} that require massive amounts of labelled data for better generalisation. 

Collecting such volumes of labelled behavioural biometric data is not practical in many real-world scenarios. For instance, collecting a significant amount of training data at registration time will inconvenience users, reduce usability, and raise privacy concerns. As a result, it is important to build learning methods that enable building deep learning models using less labelled data.}


{While collecting large volumes of labelled data for behavioural biometrics is challenging and inconvenient, collecting large volumes of unlabelled data is relatively easy. Unlabelled data can be collected while the device is in use by the user without any supervision and anonymously so that the data does not contain any personally identifiable information, eliminating threats to the user's privacy. For example, a mobile platform provider planning to build a gait-based behavioural biometric can collect unlabelled data from the motion sensors of their platform users. Therefore,  it is necessary to develop learning methods that can leverage large volumes of unlabelled data to reduce the labelled data requirement of behaviours biometrics. To this end, in this paper, we propose to use non-contrastive self-supervised learning. More specifically, we make the following contributions.}


\begin{itemize}
    \item We propose a SimSiam~\cite{chen2021exploring}-based non-contrastive learning approach and associated modifications such as shallow feature extractors and weight decay to develop label-efficient classifiers for behavioural biometrics data.
    
    \item Using two EEG-based behavioural biometrics datasets in three authentication system development scenarios, we show that the proposed non-contrastive learning approach outperforms conventional supervised learning approaches by 4\%--11\% at lower amounts of labelled data. We also show that non-contrastive learning performs comparably to a state-of-the-art multi-task learning-based baseline. 
    
    \item We conduct further experiments and provide insights into the effectiveness of different types of augmentations on the non-contrastive learning process. We also provide empirical evidence of how our modifications to SimSiam models help in the context of behavioural biometrics.
    
\end{itemize}

The rest of the paper is organised as follows. In Section~\ref{Sec:Related(updated)}, we present the related work and in Section~\ref{Sec:Methodology}, we describe the overall methodology. Next we explain the datasets and model details in Section~\ref{sec:Datasets}. Section~\ref{Sec:Results} presents the results and  Section~\ref{Sec:Analysis} presents further analysis of various model parameters' effect on performance. Finally, Section~\ref{Sec:Conclusion} discusses limitations of our work and possible extensions, and concludes the paper. 

%% file: sections/UpdatedRelated.tex
\section{Related Work}
\label{Sec:Related(updated)}

\subsection{Behavioural Biometrics}

There is a vast body of work proposing various behavioural biometric modalities. Early work involved using typing patterns and touch gestures~\cite{banerjee2012biometric,10.1145/3372224.3380901,meng2014design} while later modalities leveraged human physiology ~\cite{chauhan2017breathprint,chauhan2018breathing,sooriyaarachchi2020musicid,zhang2018mindid,zhao2020trueheart,vhaduri2017wearable}. The authentication solutions generally involve building machine learning classifiers or  signature similarity-based approaches~\cite{abuhamad2020sensor}. More recent works use deep learning methods, given their broader success in other domains~\cite{sundararajan2018deep}. 

Other works in behavioural biometrics aimed to increase the training efficiency with class incremental learning~\cite{chauhan2020contauth} or improved label efficiency using few-shot learning~\cite{solano2020few} and transfer learning~\cite{zhang2021human}. Similar efforts were also made in human activity recognition~\cite{sheng2019siamese, saeed2019multi}.

\textit{In contrast, we propose to improve the label efficiency by using non-contrastive self-supervised learning. Non-contrastive learning leverages large volumes of unlabelled data to build label-efficient classifiers. To the best of our knowledge, our work is the first to use non-contrastive learning for behavioural biometrics.}

\subsection{Self-supervised Learning (SSL)}

Self-supervised learning (SSL) refers to a broader family of methods in which a model learns representations from unlabelled data using \textit{pretext tasks}. The pretext task acts as a feature extractor for supervised learning tasks reducing the labelled data requirement. For example, in computer vision, a pretext task learning may train a model to predict whether an image is an original or an augmentation. In this way, the model learns the distinguishing features of the original image. The pretext model is then fine-tuned for a downstream task in a supervised setting with labelled data. Jing et al.~\cite{jing2020self} provide a survey of SSL methods. 

Early work closely resembling modern SSL includes Bromley et al.~\cite{bromley1993signature}, where the authors proposed the "Siamese" neural network architecture for signature verification. However, due to excessive resource requirements, SSL didn't receive much attention until their success in natural language processing. In 2013, Mikolov et al.~\cite{mikolov2013distributed} used self-supervised learning to introduce \textit{word2vec}, which paved the way to powerful generative language models such as BERT~\cite{devlin2018bert}, RoBERTa~\cite{liu2019roberta} and XLM-R~\cite{conneau2019unsupervised}.

Nonetheless, neither \textit{generative methods}~\cite{van2016conditional, dinh2016density, kingma2018glow} nor \textit{discriminative approaches}~\cite{doersch2015unsupervised, zhang2016colorful, noroozi2016unsupervised, gidaris2018unsupervised} were successful in other domains such as computer vision due to high computational complexity~\cite{chen2020simple}. In contrast, Siamese networks-based \textit{comparative methods} have shown promising results in computer vision~\cite{chen2020simple, grill2020bootstrap, caron2020unsupervised, chen2021exploring}.

The basic form of Siamese networks consists of two identical neural networks which take two views of the same input (i.e., a positive pair) and  outputs embeddings that have a low energy (or high similarity) between them. To increase the similarity of the two views, the networks learn spatial or temporal transformation invariant embeddings. Despite many successful applications of Siamese Networks, \textit{collapsing networks} (where the network converges to a trivial solution) limit their performance. 


To overcome these limitations, \textit{contrastive learning} methods ~\cite{chen2020simple,he2020momentum,dwibedi2021little,misra2020self,eldele2021time} used negatives to avoid collapsing by not only pulling positives towards each other but also by pushing apart negatives in the embedding space. An example is the SimCLR model~\cite{chen2020simple}. However, contrastive learning requires large batch sizes ~\cite{chen2020simple, eldele2021time}, support sets ~\cite{dwibedi2021little}, or memory queues ~\cite{misra2020self, chen2020improved,he2020momentum}. 

As a result, \textit{non-contrastive learning} methods, and in particular the SimSiam model ~\cite{chen2021exploring}, emerged as a viable alternative. Non-contrastive learning generally involves clustering ~\cite{caron2020unsupervised,caron2018deep}, momentum encoders ~\cite{grill2020bootstrap}, and using a cross-correlation matrix between the outputs of
two identical networks as the objective function ~\cite{zbontar2021barlow}, to address collapsing networks. These methods avoid the use of negatives to overcome the limitation of contrastive learning whereby two positive pair samples can get pushed apart in the embedding space, consequently becoming a negative pair and harming the performance of the end task~\cite{khosla2020supervised}. However, the SimSiam~\cite{chen2021exploring} outperforms other non-contrastive approaches without using complex training approaches such as momentum encoders. It emphasises the importance of stop-gradient to present an efficient and a simple solution to the collapsing networks problem. 


\subsection{SSL in Sensing and Behavioural Biometrics}

While SSL has majorly contributed to natural language processing, computer vision, and speech processing, its feasibility has been explored in sensing and mobile computing~\cite{tagliasacchi2019self}. Saeed et al.~\cite{saeed2019multi} introduced self-supervised learning for time-series sensor data by introducing augmentations that are compatible with time-series data. The authors used a multi-task SSL model to reduce the labelled training data requirement in Human Activity Recognition (HAR). Using ten labelled samples per class, the authors achieved approximately 88.8\% of the highest score reached by conventional supervised learning. SimCLR and several other contrastive and non-contrastive SSL methods also have been assessed on HAR problems ~\cite{tang2020exploring,qian2022makes}. Others such as Wright and Stewart~\cite{wright2019one} and Miller et al.~\cite{miller2021using} explored the use of traditional Siamese networks to reduce the training data requirement of  behavioural biometrics-based user authentication.

\textit{In contrast to these works, to the best of our understanding, we are the first to propose  SimSiam~\cite{chen2021exploring}-based non-contrastive learning for behavioural biometrics to reduce the labelled data requirement. Our method neither uses negatives nor requires complex training approaches such as momentum encoders to avoid collapsing. We compare our approach with baselines including traditional supervised learning, transfer learning, data augmentation, and state-of-the-art multi-task learning~\cite{saeed2019multi} and show that it can outperform supervised learning and provide comparable performance to multi-task learning at lower amounts of labelled data.}


%% file: sections/Methodology.tex
\section{Methodology}
\label{Sec:Methodology}

\begin{table*}[t!]
\centering
\begin{tabular}{p{1.5cm}p{7cm}p{7cm}}
\toprule
\textbf{Scenario} & \textbf{Research Question} & \textbf{Baseline Methods} \\\hline\hline
1 & Can non-contrastive SSL be used to leverage unlabelled data from a set of users to build a label-efficient classifier for a completely different set of users? & Supervised learning, Data augmentation,  Transfer learning, Self-supervised multi task learning, Simple Siamese network\\ \hline
2 & Can non-contrastive SSL be used to leverage unlabelled data from a given set of users to train a label-efficient classifier for user authentication? & Supervised learning, Data augmentation, Self-supervised multi task learning, Simple Siamese network \\ \hline
3 & Can non-contrastive SSL be used to leverage unlabelled data
from an initial set of users to build a label-efficient classifier for both initial set of users and a whole new set of users? & Supervised learning, Data Augmentation, Self-supervised multi-task learning, Simple Siamese network 
\\
\bottomrule
\end{tabular}
\caption{Summary of research questions and baselines}
\label{tab:baselines}
\end{table*}

\subsection{Non-contrastive Learning Approach}
\label{Sec:noncontrastive}

\begin{figure}[h]
    \centering
    \includegraphics[width=0.3\textwidth, angle=0]{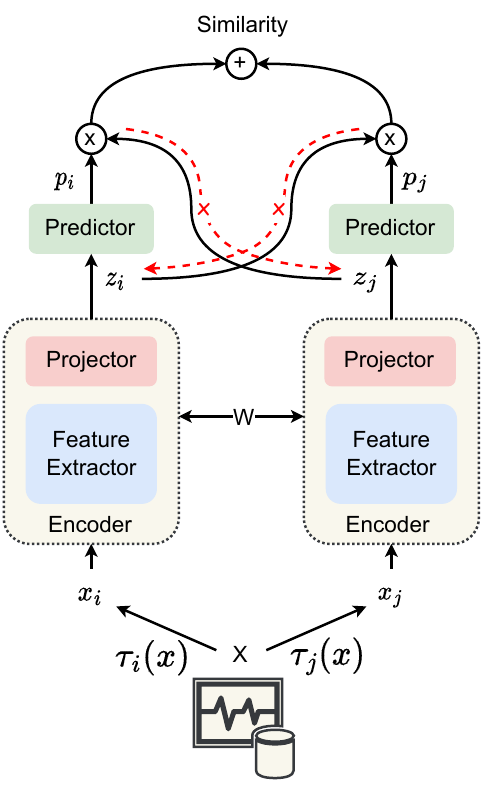}
    \caption{SimSiam Architecture}
    \label{fig:SimSiam}
\end{figure}

Our approach is based on the SimSiam architecture proposed by Chen et al. \cite{chen2021exploring}. Its architecture is a more simplified non-contrastive architecture that doesn't use negative pairs or other complex approaches to avoid collapsing. SimSiam architecture consists of two twin networks that share weights, as illustrated in Figure~\ref{fig:SimSiam}. 

The idea is to learn a good representation of inputs by solving the problem of increasing the similarity between a positive pair $(x_i, x_j)$. A positive pair consists of two randomly augmented  versions  of the same input sample $x$. That is;
$$x_i = \tau_i(x)$$
$$x_j = \tau_j(x)$$
Here $\tau$ is a function that generates a random augmentation each time it is called.
Then the two versions are encoded using the encoder network $g(x;\theta_g)$, 
$$z_i = g(x_i)$$
$$z_j = g(x_j)$$
The encoder consists of a feature extractor $g_{fe}(x;\theta_{fe})$ and a projector $g_{p}(x;\theta_{p})$. That is;
$$g(x) = g_p(g_{fe}(x))$$


The key idea of the projector is to convert the representation learnt by the feature extractor to a vector that can be used to calculate the similarity. Next, the encodings go through another \textit{predictor} network $h(x,\theta_h)$ before calculating the similarity.
$$p_i = h(z_i)$$
$$p_j = h(z_j)$$

The purpose of the predictor is to predict the average of the representation vector across all possible augmentations the network has seen~\cite{chen2021exploring}. Next, the model calculates the cosine similarity within the pairs $(p_i, z_j)$ and $(p_j, z_i)$.
$$Sim(p_i, z_j)=\dfrac{p_i.z_j}{\lVert p_i\rVert_2.\lVert z_j\rVert_2}$$
$$Sim(p_j, z_i)=\dfrac{p_j.z_i}{\lVert p_i\rVert_2.\lVert z_j\rVert_2}$$ 

Here, $\lVert.\rVert_2$ denotes the \textit{$l_2$} norm of a vector. The task of the SimSiam model is to increase the total similarity, $Sim(p_i, z_j)+Sim(p_j, z_i)$. To do that, at training time the symmetric negative cosine similarity loss as defined below is used.


$$L=-\dfrac{1}{2}Sim(p_i, stopgrad(z_j))-\dfrac{1}{2}Sim(p_j, stopgrad(z_i))$$

Note that, applying the \textit{stopgrad} operation is essential for the SimSiam architecture to work~\cite{chen2021exploring}. It considers one side of the network as constant when computing the gradients of the other side, to prevent gradients from backpropagating in that direction as shown in Figure~\ref{fig:SimSiam}. During the training process the parameters, $\theta_{fe}$, $\theta_{p}$ and $\theta_{h}$ are learnt.

\begin{figure}[h]
    \centering
    \includegraphics[width=0.45\textwidth, angle=0]{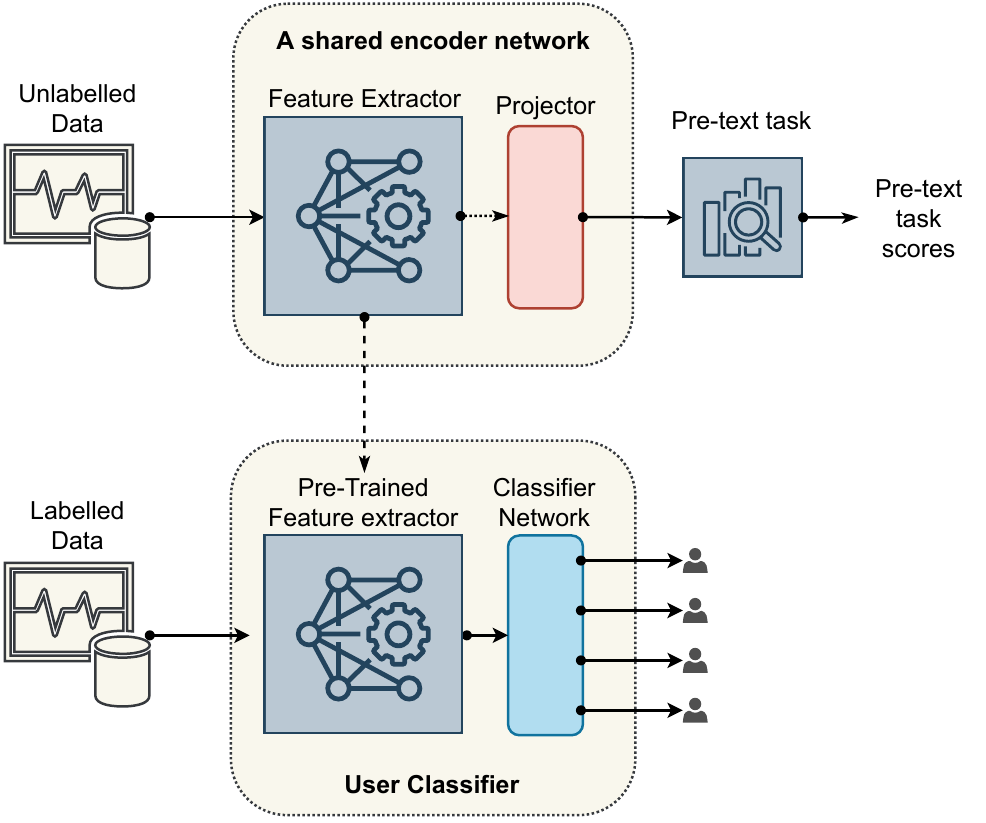}
    \caption{Using the pre-trained feature extractor to build a downstream task classifier}
    \label{fig:self-supervised}
\end{figure}

After the pre-text training of the SimSiam model, we transfer the trained feature extractor $g_{fe}(x;\theta_{fe})$ to our downstream task of building a classifier as illustrated in Figure~\ref{fig:self-supervised}.


We introduce two modifications to make SimSiam architecture work for time series behavioural biometrics data and further improve its performance. They are based on the hypothesis that easier self-supervision tasks lead to learning useless features and such features do not hold any value for subsequent downstream tasks. Therefore, it is important to make the self-supervised learning part more challenging so that robust features are learned during pre-text training. 

\begin{itemize}
    \item {\bf Shallow feature extractor networks - }The original SimSiam architecture was designed for image data. Time series data of behavioural biometrics are less complex compared to images and as such, to avoid the model over-fitting on pre-text the task, we use shallow feature extractor networks. A shallow feature extractor makes the learning task more difficult and as a result, allows the building of better feature extractors.
    
    \item {\bf Weight decay - }Using weight decay to prevent over-fitting is common in any machine learning application. We add high weight regularisation to the feature extractor network to avoid overfitting and make the pre-text training more challenging.
    
\end{itemize}

Later, in Section~\ref{SubSec:Modifications} we provide an analysis of how the performance improves with our modifications. During our experiments, we also came across another important finding about the predictor network. We found that a deeper predictor network can help improve the non-contrastive SSL process. We provide further analysis and an explanation as to why it is happening in Section~\ref{SubSec:Predictor}.




\subsection{Authentication Scenarios}
\label{SubSec:Scenarios}

We conduct experiments to demonstrate the effectiveness of non-contrastive SSL in behavioural biometrics under three scenarios that are commonplace in authentication settings. 

\begin{itemize}

    \item {\bf Scenario 1 -} This scenario represents what is usually encountered by anyone who is developing a large-scale behavioural biometrics solution. That is, it is possible to collect large volumes of unlabelled data. For example, mobile OS providers can collect unlabelled data streams such as touch patterns and gait patterns from a large number of users, in compliance with privacy regulations. However, only a limited amount of labelled data can be collected from a known set of users due to usability and privacy constraints, either in-house or explicitly obtaining customer consent. \\\vspace{-2mm}
    
    More specifically, for $N_1$ users, a large volume of unlabelled data $X_{U_1}$ is available. For different $N_2$ users, only a limited amount of  labelled data $(X_{L_{2_{min}}},Y_2)$ is available. The task is to build a classifier $f_{SSL}(x;\theta)$ to identify the user
    $y \in \{1,...,N_2\}$ given the input $x$,  by using both unlabelled data $X_{U_1}$ and limited labelled data $(X_{L_{2_{min}}},Y_2)$. Here $|X_U| \gg |X_{L_{min}}|$. \\\vspace{-2mm}
    
    Here, we use the unlabelled data from $N_1$ users $X_{U_1}$ to pre-train  a SimSiam-base feature extractor. Next, we train a classifier network on top of the pre-trained feature extractor for $N_2$ users. We use the labelled data $(X_{L_{2_{min}}},Y_2)$ to train the classifier. We fine-tune the learnt weights of the feature extractor while training the classifier network. The concatenated fine-tuned feature extractor and the classifier network creates the final classifier $f_{SSL}(x;\theta)$ ({\bf cf.} Figure~\ref{fig:self-supervised}). \\\vspace{-2mm}

    \item {\bf Scenario 2 -} This scenario is similar to Scenario 1. However here an organisation is trying to build an in-house authentication system. As a result, again there is a large volume of unlabelled data and a limited about of labelled data, but for the same set of users in contrast to Scenario 1. \\\vspace{-2mm}
    
    That is, for $N$ users, only a limited amount of labelled data, $(X_{L_{min}},Y)$ is available while a large volume of unlabelled data $X_{U}$ is available. The task is to build a classifier, $f_{SSL}(x;\theta)$ to correctly identify the user $y \in \{1,...,N\}$ given an input data sample, $x$, by using the limited labelled data, $X_{L_{min}}$ and unlabelled data $X_U$. Again here $|X_U| \gg |X_{L_{min}}|$. \\\vspace{-2mm}
    
    Similar to Scenario 1, here also  we first pre-train a SimSiam-based feature extractor using unlabelled data $X_U$ and build a classifier network on top of the feature extractor for $N$ users using available limited labelled data, $(X_{L_{min}},Y)$. During classifier training, feature extractor fine-tuning happens the same as Scenario 1. The concatenated fine-tuned feature extractor and the classifier network create the final classifier $f_{SSL}(x;\theta)$. \\\vspace{-2mm}
    
    
    
    
    

    \item {\bf Scenario 3 -} This is an incremental step from Scenario 2, where the organisation has collected labelled and unlabelled data for building the  authentication system but has additional new users for whom only a limited amount of labelled data is available. \\\vspace{-2mm}
    
    That is, for $N_1$ users, a limited amount of labelled data, $(X_{L_{1_{min}}},Y_1)$ and a large volume of unlabelled data, $X_{U_1}$ is available. For different $N_2$ users, only a limited amount of labelled data $(X_{L_{2_{min}}},Y_2)$ is available. The task is to build a classifier, $f_{SSL}(x;\theta)$ to correctly identify the user $y \in \{1,...,(N_1+N_2)\}$ from the combined set, given an input data sample, by leveraging both unlabelled and labelled data. \\\vspace{-2mm}
    
    Here, we first use unlabelled data from $N_1$ users $X_{U_1}$ to pre-train the feature extractor with the SimSiam architecture. Then, we build a classifier on top of the pre-learned feature extractor, for the combined set of $N_1+N_2$ users using both labelled datasets $(X_{L_{1_{min}}},Y_1)$ and $(X_{L_{2_{min}}},Y_2)$. Similar to previous scenarios, we fine-tune the learnt weights of the feature extractor while training the classifier network. The concatenated fine-tuned feature extractor and the classifier network create the final classifier $f_{SSL}(x;\theta)$. \\\vspace{-2mm}

\end{itemize}



\subsection{Baseline Methods}

We compare our non-contrastive SSL approach with multiple baselines. Below we provide a general overview of the different baselines we use. However, we highlight that not all baselines apply for all three scenarios. 

\begin{enumerate}
    \item {\bf Supervised learning} - We train a 1D CNN based on available labelled data. For example, for Scenario 1, we leverage the available limited labelled data, $X_{L_{min}}$, and train $f_{S}(x;\theta)$. We do all the required hyperparameter tuning such as finding optimal convolution kernel sizes, number of convolutional filters in a layer, depth of the network, learning rates, and weight regularisation constants to ensure that we leverage the full capability of the supervised learning approach. 
    
    \item {\bf Data augmentation} - Data augmentation is a default step in any deep neural network training process. It helps to increase the generalisability of the model as well as learn from limited labelled data to some extent. In this baseline, we augment available limited labelled training data using two methods, scaling, and noise addition and continue to train a supervised learning classifier, as usual, using both limited labelled data and augmented data. 
    
    \item {\bf Multi-task self-supervised learning (MTSSL)} - This is a self-supervised learning baseline, which can leverage unlabelled data compared to the above supervised learning approach. As a result, it is a closer baseline to our approach of non-contrastive SSL. Here, we first train a multi-task model with a common feature extractor using unlabelled data available in a given scenario. The common feature extractor is connected to several heads, each having a dedicated discriminative task. Each head is a binary classifier learning to discriminate whether an assigned augmentation is applied to a sample or not. Next, we build a classifier on top of the pre-trained feature extractor by adding extra fully connected layers. 
    
    \item {\bf Transfer learning} - Transfer learning is the most common approach to handling the lack of labelled data. Here, the deep neural network trained using labelled data is leveraged as a feature extractor to facilitate adding new users to an existing behavioural biometric system or building a new behavioural biometric system from less labelled data. During the transfer learning phase, several new layers are added to the previous feature extractor, corresponding to the new classification task. The entire model is then fine-tuned with the available limited training data. 

    \item {\bf Transfer learning with data augmentation} - Here, we do transfer learning together with data augmentation.
    
\end{enumerate}

In Table~\ref{tab:baselines} we summarise the research questions associated with the three scenarios  and the baselines used in each scenario. We give further details of the implementations aspects of the baselines methods in Section~\ref{Sec:ModelImplementation}.


\subsection{Performance Metrics}

To measure the performance of trained non-contrastive SSL-based user authentication systems and compare them against other baselines we use \textbf{Cohen’s Kappa coefficient} similar to~\cite{saeed2019multi}. Usually, accuracy is the most commonly used metric to evaluate the performance in a multi-class setting. Accuracy measures the agreement between two raters, here raters being the true label and predicted label.

$$\text{Accuracy} = \dfrac{\text{No. of correct predictions}}{\text{No. of total predictions}}$$



 However, accuracy can be misleading in some occasions especially if the trained model is biased to predict one class more accurately and another class less accurately. In contrast \textbf{Cohen’s Kappa coefficient} measures the agreement between two raters but discount the effect of agreement by chance. That is, 

$$\text{Kappa Score} = \dfrac{P_{o}-P_{e}}{1-P_{e}}$$\\
$P_{o} = \text{Probability of agreement (Accuracy)}$\\
$P_{e} = \text{Probability of agreement by chance}$\\\vspace{-3mm}

That is,

$$P_{o} = \dfrac{\text{No. of correct predictions}}{\text{No. of total predictions}}$$
$$P_{e} = \sum_i^N \dfrac{n_{true}^i}{n_{total}}\times \dfrac{n_{pred}^i}{n_{total}}$$\\
$N = \text{Number of Classes}$\\
$n_{total} = \text{No. of total predictions}$\\
$n_{true}^i = \text{No. of true labels of the $i^{th}$ class}$\\
$n_{pred}^i = \text{No. of predicted labels of the $i^{th}$ class}$ \\


The highest possible Kappa Score is 1 indicating the best performance and it can have negative values at worse performances. Overall, we progressively increase the amount of labelled data available and compare the performance of different methods using the Kappa Score. We further discuss this in Section~\ref{Sec:Results}.


\section{Datasets and Models}
\label{sec:Datasets}

\subsection{Datasets}
To demonstrate the effectiveness of non-contrastive learning in
behavioural biometrics, we use two datasets; {\bf MusicID}~\cite{sooriyaarachchi2020musicid} and {\bf MMI}~\cite{schalk2004general}. Both of these datasets are
EEG datasets and have been used in behavioural biometric settings
before. Note that, for the rest of the paper, a \textit{session} refers to a single experiment of recording sensor readings in one sitting for one user.

\begin{itemize}
    \item {\bf MusicID - }This dataset consists of brainwave data collected from 20 volunteers while they performed two tasks~\cite{sooriyaarachchi2020musicid}; listening to a popular English song and listening to the individual’s favourite song. The participants wore a Muse brain sensing headset while listening to the music and kept their eyes closed. The dataset experiment was approved by the host institution's Human Research Ethics Committee as mentioned in the original work. The duration of each task in a single session was 150s and the headset records samples at a rate of 2Hz, resulting a total of 300 readings per participant, per session. Data was collected from each participant over multiple sessions with the number of  sessions per user varying between 12-30 (considering both the same song and favourite song sessions together). Each headset recording contains 24 readings; absolute brainwave values of \textit{alpha}, \textit{beta}, \textit{theta}, \textit{delta}, \textit{gamma}, and \textit{raw EEG} from the standard 4-channels of the Muse headset. That is, a single reading in this dataset is a 24 dimensional vector, $x_i \in \mathbb{R}^{24}$. We use 30 readings (i.e., 15 seconds of data) as a single input sample to the model. Consequently, one input to our model is $x \in \mathbb{R}^{30 \times 24}$. \\
    \vspace{-2mm}
    
    \item {\bf MMI - } This is a publicly available dataset known as \textit{eegmmidb} (EEG {\bf M}otor {\bf M}ovement/{\bf I}magery database). It comprises of EEG signals obtained from 109 participants using the BCI2000 system~\cite{schalk2004general}. Separate experiments were conducted where each participant carried out four different tasks, each for a two-minute duration and repeated three times. The tasks involved different combinations of opening and closing fists or feet based on the location of a certain target shown on a screen. Each participant also performed two one-minute baseline runs. The data was obtained in EDF+ format – consisting of 64 EEG signals and an annotation channel where the annotation channel indicates the participant's activity. We randomly selected eight channels; 3, 12, 13, 18, 50, 60, 61, and 64 to reduce the memory requirements. For each channel, we filter the raw signal to get the \textit{alpha}, \textit{beta}, \textit{theta}, and \textit{delta} components and use the raw signal and the four filtered components as input features to our models. That is, a single reading $x_i \in \mathbb{R}^{40}$. The sessions are recorded with a $160Hz$ sampling rate. We use $800ms$ of readings making our inputs to the models $x \in \mathbb{R}^{128 \times 40}$.\\\vspace{-2mm}
    
    Even though this dataset has been used for user authentication~\cite{zhang2018mindid}, the previous work only use a portion of the dataset, \textit{EEG-S}, which only contains data from eight users. However, we, on the other hand, incorporate all 109 users in our experiments. Due to the high number of users and the fact that this dataset was not collected with authentication as a target application, the maximum kappa scores we could achieve for user authentication in the \textbf{MMI} dataset is less than the \textbf{MusicID} dataset.    
    
\end{itemize}

\subsection{Dataset Splits}
\label{Sec:DatasetSplit}

To emulate the three scenarios described in Section~\ref{SubSec:Scenarios}, we split each dataset into two parts; {\bf Dataset 1} and {\bf Dataset 2}. \textbf{Dataset 1} contains data of approximately 1/3 of the users of the total dataset and \textbf{Dataset 2} contains the rest. We use these two datasets in different ways in the three scenarios. Note that in the following description, "\textit{labelled data} of \textbf{Dataset 1}/\textbf{Dataset 2}" refers to $(x, y)$ pairs coming from a dataset while "\textit{unlabelled data} of \textbf{Dataset 1}/\textbf{Dataset 2}" refers to only $(x,)$ values coming from a dataset, ignoring the user ID labels.

In Scenario 1, we use \textbf{Dataset 1} as the unlabelled data $X_{U_1}$ coming from $N_1$ users. We use labelled data from \textbf{Dataset 2} as the labelled data $(X_{L_{2_{min}}},Y_2)$ coming from $N_2$ users who will be part of the authentication system. We progressively increase the amount of labelled data in $(X_{L_{2_{min}}},Y_2)$ to assess the performance of our method and other methods as presented in Section~\ref{SubSec:Scenario1}.


Since in Scenario 2 both the labelled and unlabelled data are associated with the same set of users, we use only \textbf{Dataset 2}. In Scenario 3 we use \textbf{Dataset 1} as the dataset coming from the initial users $N_1$ who provide both unlabelled $X_{U_1}$ and labelled data $(X_{L_{1_{min}}},Y_1)$. We use labelled data from \textbf{Dataset 2} as the labelled dataset $(X_{L_{2_{min}}},Y_2)$ coming from the new set of users $N_2$.

We use dedicated sessions for validation and testing. Reading training data is done with a 50 percent overlapping window  similar to \cite{sooriyaarachchi2020musicid}. We do not use overlapped windows when reading validation and test data since it will artificially increase the performance metrics. We summarise the two datasets and the splits in Table~\ref{tab:datasets}.


\begin{table}[t]
\centering
\smaller
\begin{tabular}{p{3.2cm}p{1.8cm}p{1.8cm}}
\toprule
\textbf{Dataset} & \textbf{MusicID~\cite{sooriyaarachchi2020musicid}} & \textbf{MMI~\cite{schalk2004general}} \\\hline\hline

No. of users & 20 & 109 \\ 
Sessions/user & 12-30 & 14 \\
Samples/user & 68-170 & 3,242-4,119 \\
\vspace{1mm}\\
\midrule

{\bf Split - Dataset 1} &   &  \\ \bottomrule
No. of users & 6 & 36 \\\vspace{-1mm}

{\it Unlabelled Data} &   &  \\\hline
\quad Training sessions/user & 8-10 & 6 \\
\quad Training samples/user & 120-150 & 1,524-1,546 \\
\vspace{1mm}

{\it Labelled Data} &   &  \\\hline
\quad Training sessions/user & 8-10 & 2 \\
\quad Training samples/user & 120-150 & 612-622\\\hline

\quad Validation sessions/user & 8-10 & 1 \\
\quad Validation samples/user & 8-10 & 153-156\\\hline

\quad Testing sessions/user & 8-10 & 1 \\
\quad Testing samples/user & 8-10 & 153-156\\
\vspace{1mm}\\
\midrule
{\bf Split - Dataset 2} &   &  \\ \bottomrule
No. of users & 14 & 73 \\ \vspace{-1mm}

{\it Unlabelled Data} &   &  \\\hline
\quad Training sessions/user & 4-6 & 6 \\
\quad Training samples/user & 60-90 & 1282-1671 \\
\vspace{1mm}

{\it Labelled Data} &   &  \\\hline
\quad Training sessions/user & 4-6 & 2 \\
\quad Training samples/user & 60-90 & 490-757\\\hline

\quad Validation sessions/user & 4-6 & 1 \\
\quad Validation samples/user & 4-6 & 123-156\\\hline

\quad Testing sessions/user & 4-6 & 1 \\
\quad Testing samples/user & 4-6 & 123-156\\\bottomrule

\end{tabular}
\caption{Dataset and data split summary}
\label{tab:datasets}
\end{table}


\subsection{Deep Learning Models and Training}
\label{Sec:ModelImplementation}

\subsubsection{Model architectures}

\begin{figure}[h]
    \centering
    \includegraphics[width=0.45\textwidth, angle=0]{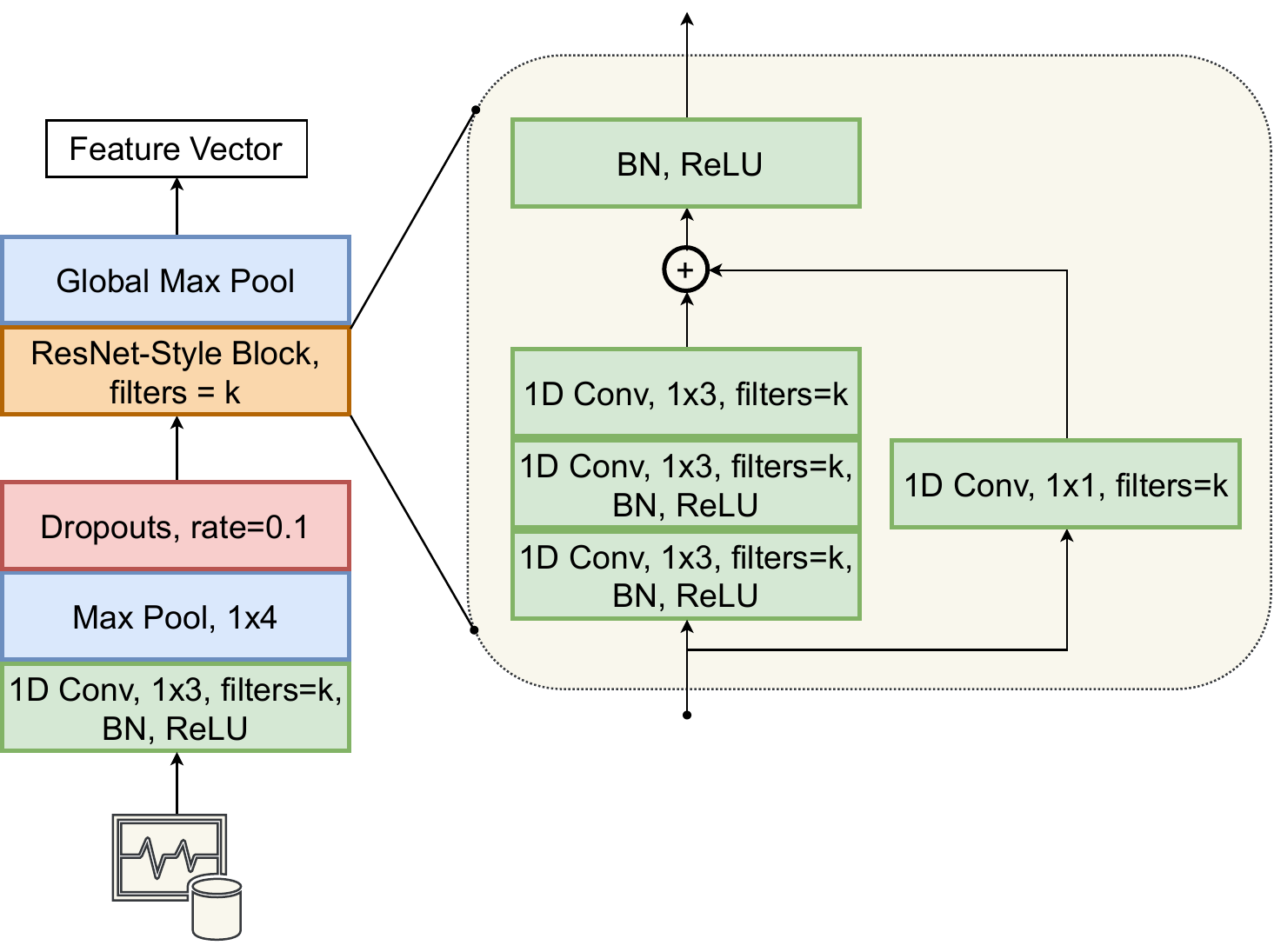}
    \caption{Feature extractor architecture}
    \label{fig:feature}
\end{figure}

We use the same 1D ResNet architecture illustrated in Figure~\ref{fig:feature} as the feature extractor across all datasets, experiments, and models. The reason behind this choice is it allows to effectively compare the learning methods by minimising the effect of the network architecture. The only difference between the models in the two datasets is the number of convolutional filters, which is represented by $k$ in Figure ~\ref{fig:feature}. For example, for the \textbf{MuiscID} dataset we used $k=(128,256)$, meaning the first convolutional layer has 128 filters and the ResNet block has 256 filters. For the \textbf{MMI} dataset it was $k=(48, 96)$. 



\begin{figure}[h]
    \centering
    \includegraphics[width=0.35\textwidth, angle=0]{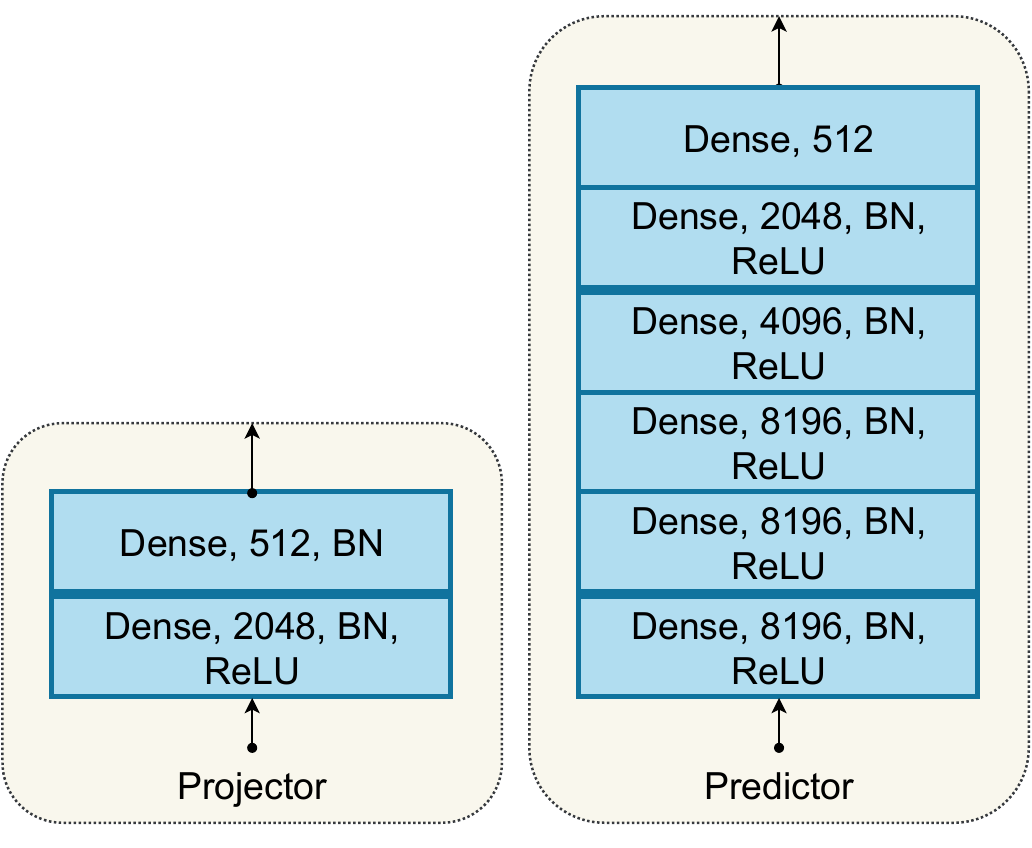}
    \caption{Projector and Predictor Networks}
    \label{fig:proj_pred}
\end{figure}

We used MLP architectures illustrated in Figure~\ref{fig:proj_pred} for the \textit{Projector} and \textit{Predictor} networks of the SimSiam model. Finally, the classifier networks were also MLPs, consisting of two hidden Dense layers with the dimensions of 256, 64. The number of neurons in the final layer was equal to the number of users of the corresponding task. We used \textit{ReLU} activation function for all the MLP layers except for the final layer, which used the \textit{softmax} activation.

\subsubsection{Input Transformations for SimSiam and MTSSL}

For the \textbf{MusicID} dataset we used \textit{Random Scaling} and \textit{Jitter} to create the two augmented versions inputs required for SimSiam model training. In Random Scaling, we multiply each channel of an input sample with a randomly generated variable having a normal distribution $\sim N(1,0.65)$. In Jitter, we generate a noise matrix having the same dimensions as the input. The values of the noise matrix are sampled from a normal distribution $\sim N(0,0.8)$ and added to the original input sample. We selected the values for both the variances through experiments.

Similarly, for the \textbf{MMI} dataset we use \textit{Random Scaling} and \textit{Flipping} as the augmentations. In Flipping, we reverse the time dimension of the input. We made these choices based on some early experiments where we tried various data transformation methods individually and in combination, as explained later in Section~\ref{Sec:Transformations}.


According to the findings by Saeed et al.~\cite{saeed2019multi}, using multiple augmentations for multi-task learning leads to higher self-supervised learning performances. Therefore, after a similar pre-experiment as earlier, for the \textbf{MusicID} dataset we use \textit{JItter, Random Scaling, Magnitude Warping, Random Sampling, Flipping, Data Dropping, Time Warping, Negation, Channel Shuffling, and Permutations} as transformations in our multi-task learning baseline. For the \textbf{MMI} dataset we use only four augmentations; \textit{Random Scaling, Magnitude Warping, Time Warping, and Negation.} This is because the dataset’s larger size requires high computing memory. In fact this is a limitation of multi-task learning compared to the SimSiam approach as we discuss in Section~\ref{Sec:Conclusion}. We summarise all the transformations we used to generate the results in Section~\ref{Sec:Results} and Section~\ref{Sec:Analysis} in Table~\ref{tab:transformations}.

Note that here we use the same multi-headed architecture as Saeed et al.~\cite{saeed2019multi} where each head of the multi-task model tries to discriminate whether an assigned augmentation is applied to a sample or not. For example, the first head tries to discriminate whether Gaussian noise is added to a sample or not, and the second head tries to discriminate whether a sample is scaled or not.

\begin{table*}[t]
\centering
\smaller
\begin{tabular}{p{0.5cm}p{3cm}p{12cm}}
\toprule
 & \textbf{Augmentation} & \textbf{Description}\\\midrule
1 & Jitter & Adding random noise to a sample\\ \hline
2 & Random Scaling & Scaling each channel of the input with a randomly generated constant\\ \hline
3 & Magnitude Warping & Random element-wise scaling with a smooth transition along time dimension\\ \hline
4 & Time Warping & Stretches the data across time dimension. New samples are generated using interpolation (based on the entire sample) to stretch the time dimension\\ \hline
5 & Flipping & Reversing the time dimension of a sample\\ \hline
6 & Data Dropping & Making parts of the input zero\\ \hline
7 & Random Sampling & This is similar to \textit{Time Warping}, with only a subset of sample is used for interpolation \\ \hline
8 & Permutations & Randomly slicing and swapping values across the time dimension, within moving time window s\\ \hline
9 & Negation & Multiplying the sample by -1\\ \hline
10 & Channel Shuffling & Randomly exchanging the order of the input channels\\\bottomrule

\end{tabular}
\caption{Summary of transformations/augmentations}
\label{tab:transformations}
\end{table*}

\subsubsection{Training Process}

We use Adam as the optimiser with an exponentially decaying learning rate for both pre-training and classifier training. Initial learning rates are $0.00003$ and $0.01$ for pre-training and classifier training, respectively. As explained in Section~\ref{Sec:noncontrastive}, we use $l_2\;norm$ weight regularisation with a regularisation factor of $\lambda = 0.01$ for the feature extractor networks of self-supervised learning methods. We train self-supervised models up to 30 and 10 epochs, for \textbf{MusicID} and \textbf{MMI} datasets, respectively. We train the final classifiers up to 30 epochs in both datasets. All the time we use \textit{early stopping} to prevent over-fitting.

\textit{We are planning to release all of our codes and data splits publicly upon the acceptance of the paper for reproducibility of our results and to stem further research in the area.}

%% file: sections/Results.tex
\section{Results}
\label{Sec:Results}

Next, we present the results for the three scenarios. For each scenario, we progressively increase the amount of available labelled data (i.e., the percentage of labelled samples per user) and compare the non-contrastive SSL approach with other baseline methods. Ideally, we expect to reach a high kappa score (close to one) using as few data samples as possible. Each result we report is the average of ten experiment runs to avoid any biases in weight initialisation and data splits.




\subsection{Scenario 1}
\label{SubSec:Scenario1}

\noindent{{\bf RQ1} - \textit{Can non-contrastive SSL be used to leverage unlabelled data from a set of users to build a label-efficient classifier for a completely different set of users?}} \\\vspace{-2mm}




\begin{figure}[t]
\centering \vspace{-3mm}
\subfloat[MusicID]{\label{fig:MusicID_1}\includegraphics[trim=0cm 0cm 0cm 0cm, clip=true,scale=0.16]{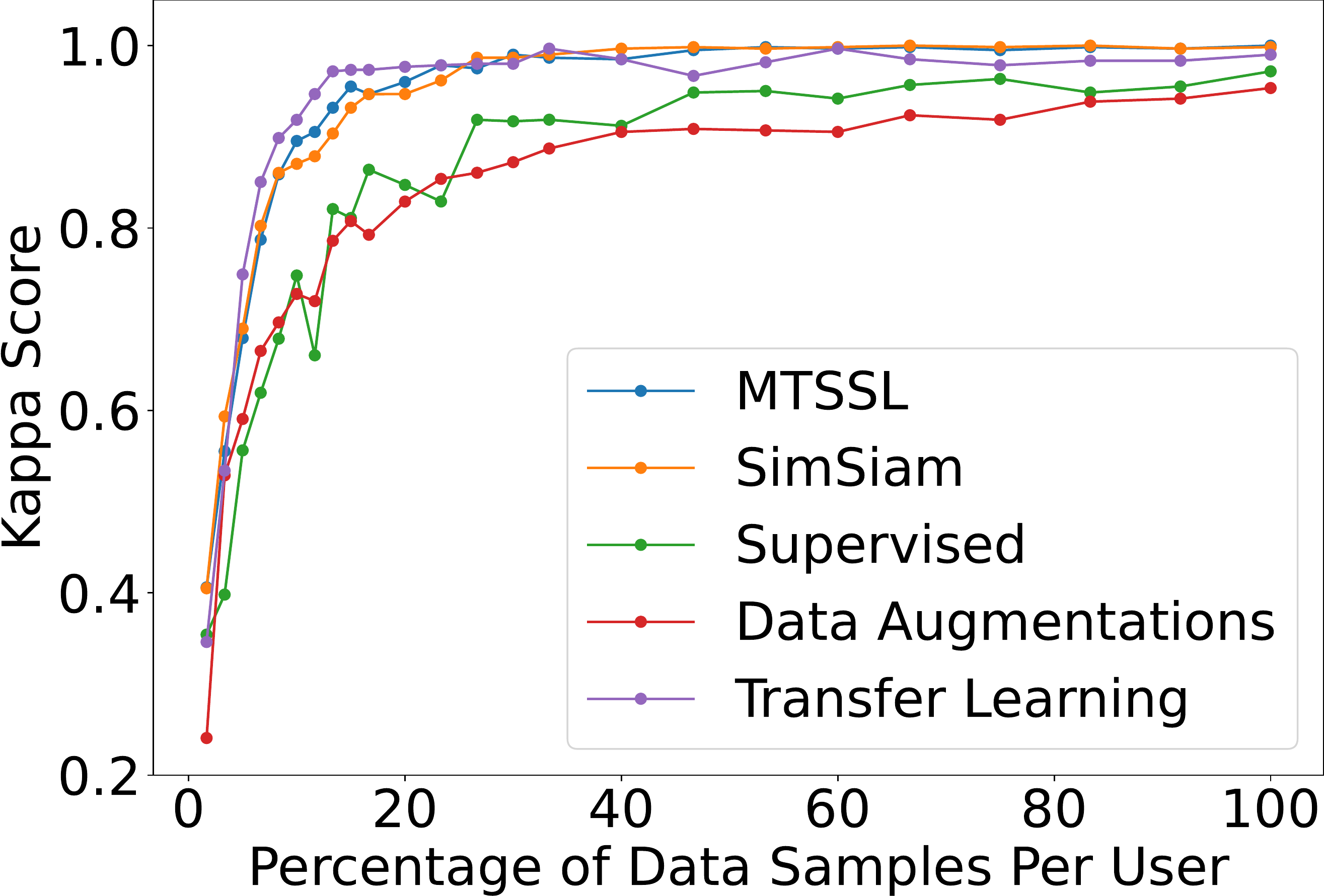}}
\subfloat[MMI]{\label{fig:MMI_1}\includegraphics[trim=0cm 0cm 0cm 0cm, clip=true,scale=0.16]{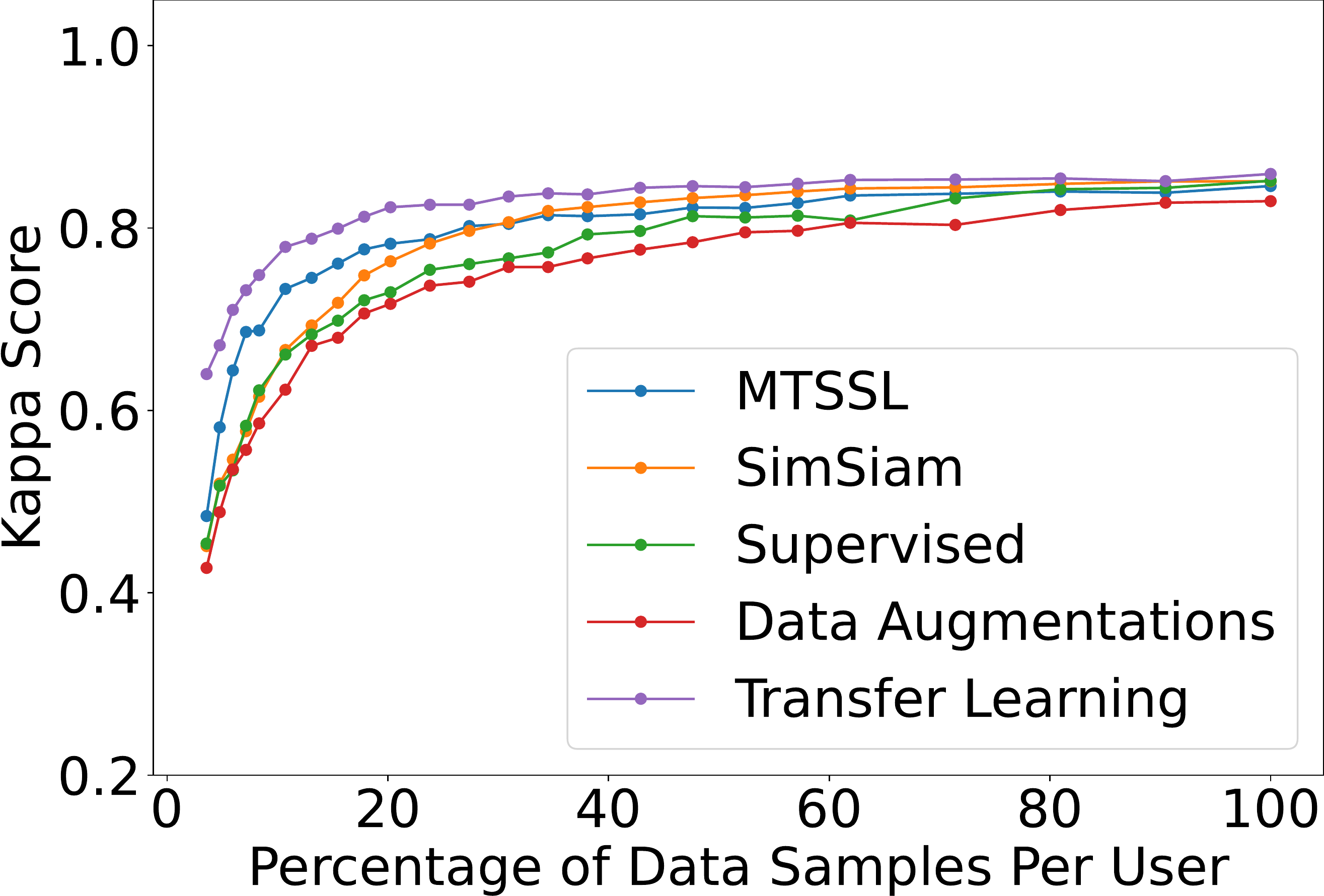}}
\caption{Performance results for Scenario 1}\vspace{-4mm}
\label{fig:scen_1}
\end{figure}

In Figure~\ref{fig:MusicID_1} and Figure~\ref{fig:MMI_1} we show the performance of our SimSiam method and various other baselines for \textbf{MusicID} and \textbf{MMI} datasets, respectively. The overall kappa score is low for the \textbf{MMI} dataset (approximately between 80\% - 85\%) because it is a much noisier dataset not necessarily designed for user authentication ({\bf cf.} Section~\ref{sec:Datasets}). 

Both the figures show that conventional supervised learning and data augmentation do not perform well because there is not enough labelled data to train them for the second set of users. For both the datasets, transfer learning works best (i.e., shows a high kappa score at a given percentage of data samples per user) followed by multi-task learning and our SimSiam approach. On the \textbf{MusicID} dataset, SimSiam's performance is competitive with multi-task learning (MTSSL), and on the \textbf{MMI} dataset, up to until 20\% of data samples per user, SimSiam is slightly worse, but catches up with multi task learning afterwards.

The high performance of transfer learning can be attributed to the use of labelled data from \textbf{Dataset 1} during the pre-training phase of transfer learning. Here, SimSiam SSL still could compete with the performance of the transfer learning without using any labelled data from \textbf{Dataset 1} which indicates the its capability in extracting features from unlabelled data.

Overall, our results show that non-contrastive SSL can indeed learn generic features from unlabelled data of a set of users to build a classifier for a totally different set of users using less labelled data. For instance, for the \textbf{MusicID} dataset, between 20\% to 40\% samples per user, the average kappa scores for SimSiam was 0.978. Within the same percentages of labelled samples per user, the average kappa scores of supervised learning and data augmentation were 0.879. As a percentage improvement this corresponds to 11\% improvement over the baselines. The corresponding percentage increase of the \textbf{MMI} dataset was approximately 4\%. Finally, though multi-task SSL performs slightly better than SimSiam it is computationally expensive as we explain in Section~\ref{Sec:Conclusion}.


\subsection{Scenario 2}
\label{SubSec:Scenario2}

\noindent{{\bf RQ2} - \textit{Can non-contrastive SSL be used to leverage large volumes of unlabelled data from a given set of users to train a label-efficient classifier for user authentication?}} \\\vspace{-2mm}

\begin{figure}[t]
\centering \vspace{-3mm}
\subfloat[MusicID]{\label{fig:MusicID_2}\includegraphics[trim=0cm 0cm 0cm 0cm, clip=true,scale=0.16]{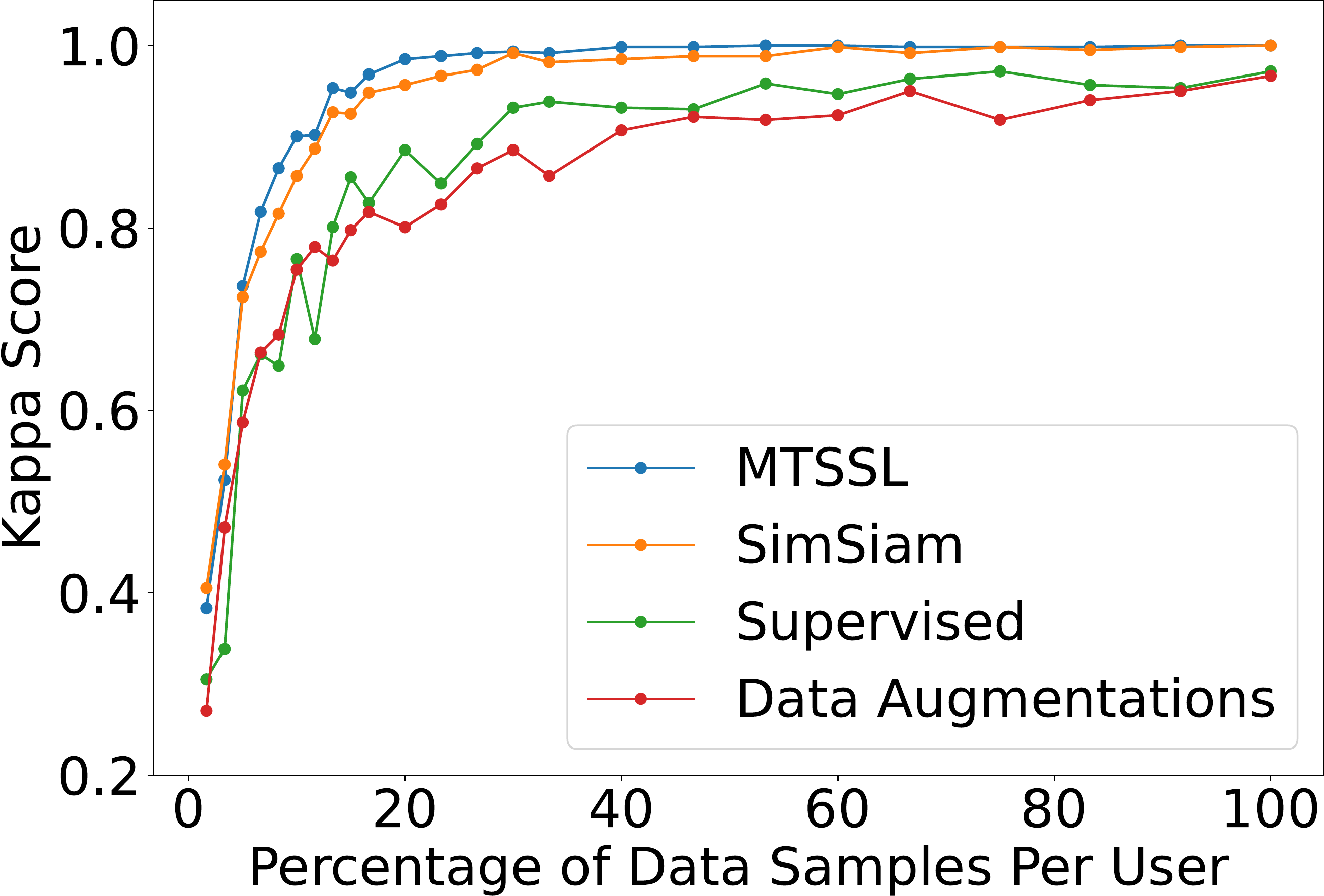}}
\subfloat[MMI]{\label{fig:MMI_2}\includegraphics[trim=0cm 0cm 0cm 0cm, clip=true,scale=0.16]{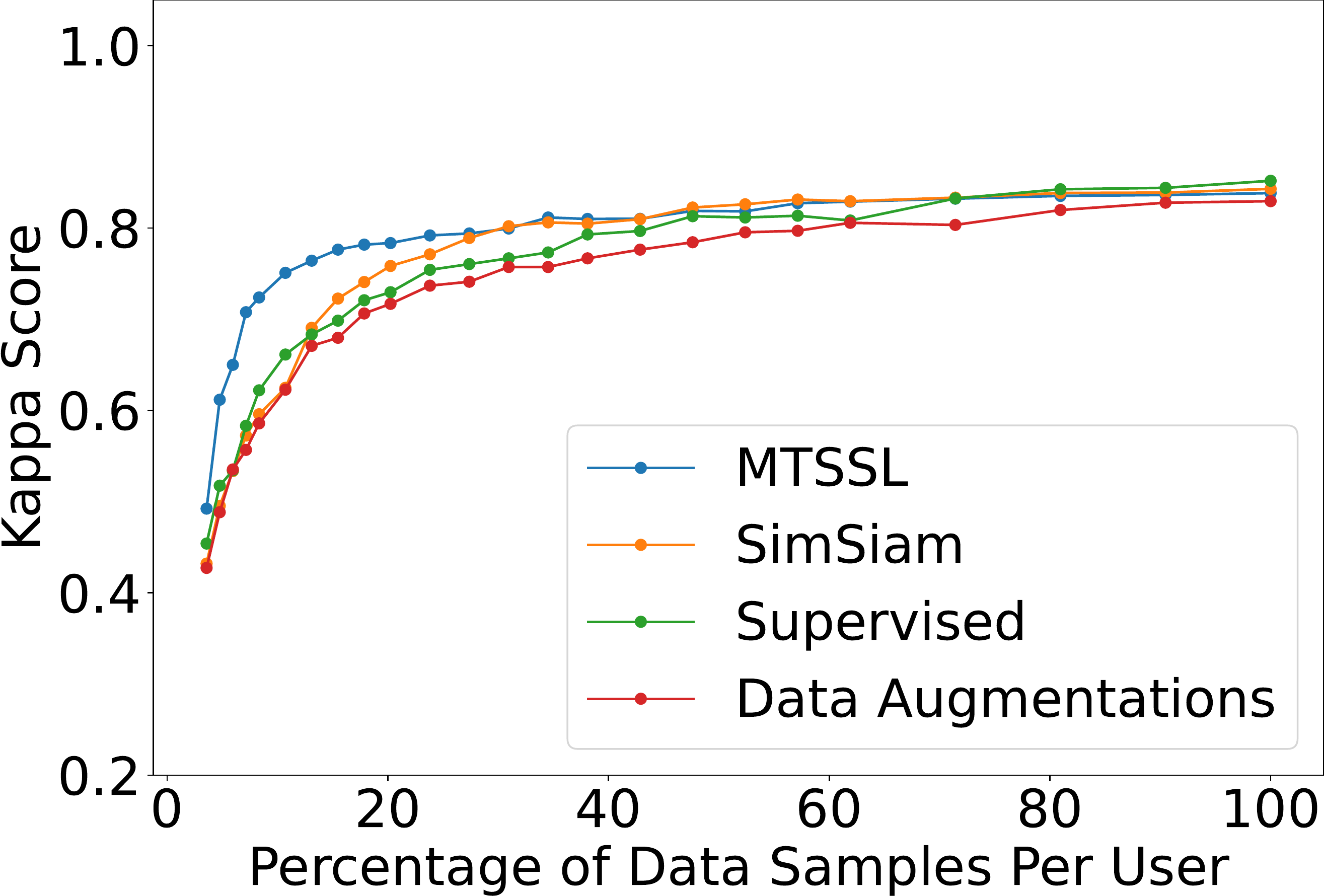}}
\caption{Performance results for Scenario 2}\vspace{-4mm}
\label{fig:scen_2}
\end{figure}

In Figure~\ref{fig:MusicID_2} and Figure~\ref{fig:MMI_2} we compare the results for Scenario 2 for the two datasets. At lower amounts of labelled data two self-supervised learning approaches; SimSiam and multi-task learning (MTSSL) are performing better compared to supervised learning approaches. For example, for the \textbf{MusicID} dataset, when only 20\% of labelled samples are used, both SimSiam and MTSSL have average kappa scores of 0.956 and 0.985, respectively while supervised learning and data augmentation approaches only result in kappa scores of 0.885, and 0.800. This difference drops in the \textbf{MMI} dataset, which can be expected since it is a much larger dataset compared to the \textbf{MusicID} dataset and when 20\% of labelled samples are used, it is sufficient to train the supervised learning classifier.  


Finally, similar to Scenario 1, it is noticeable that between the two self-supervised learning methods, the multi-task approach shows better performance than the non-contrastive SimSiam approach and the difference is more visible in \textbf{MMI} dataset compared to the \textbf{MusicID} dataset. Nonetheless, our results show that unlabelled data can be leveraged using non-contrastive SSL to build more label-efficient classifiers. For instance, the average performance improvements of SimSiam over supervised and data augmentation approaches were 10\% and 4\% for the \textbf{MusicID} and \textbf{MMI} datasets, respectively. \vspace{-1mm}



\subsection{Scenario 3}
\label{SubSec:Scenario3}

\noindent{{\bf RQ3} - \textit{Can non-contrastive SSL be used to leverage unlabelled data
from an initial set of users to build a label-efficient classifier for both the initial set of users and a whole new set of users?}} \\\vspace{-2mm}

\begin{figure}[t]
\centering \vspace{-3mm}
\subfloat[MusicID]{\label{fig:MusicID_3}\includegraphics[trim=0cm 0cm 0cm 0cm, clip=true,scale=0.16]{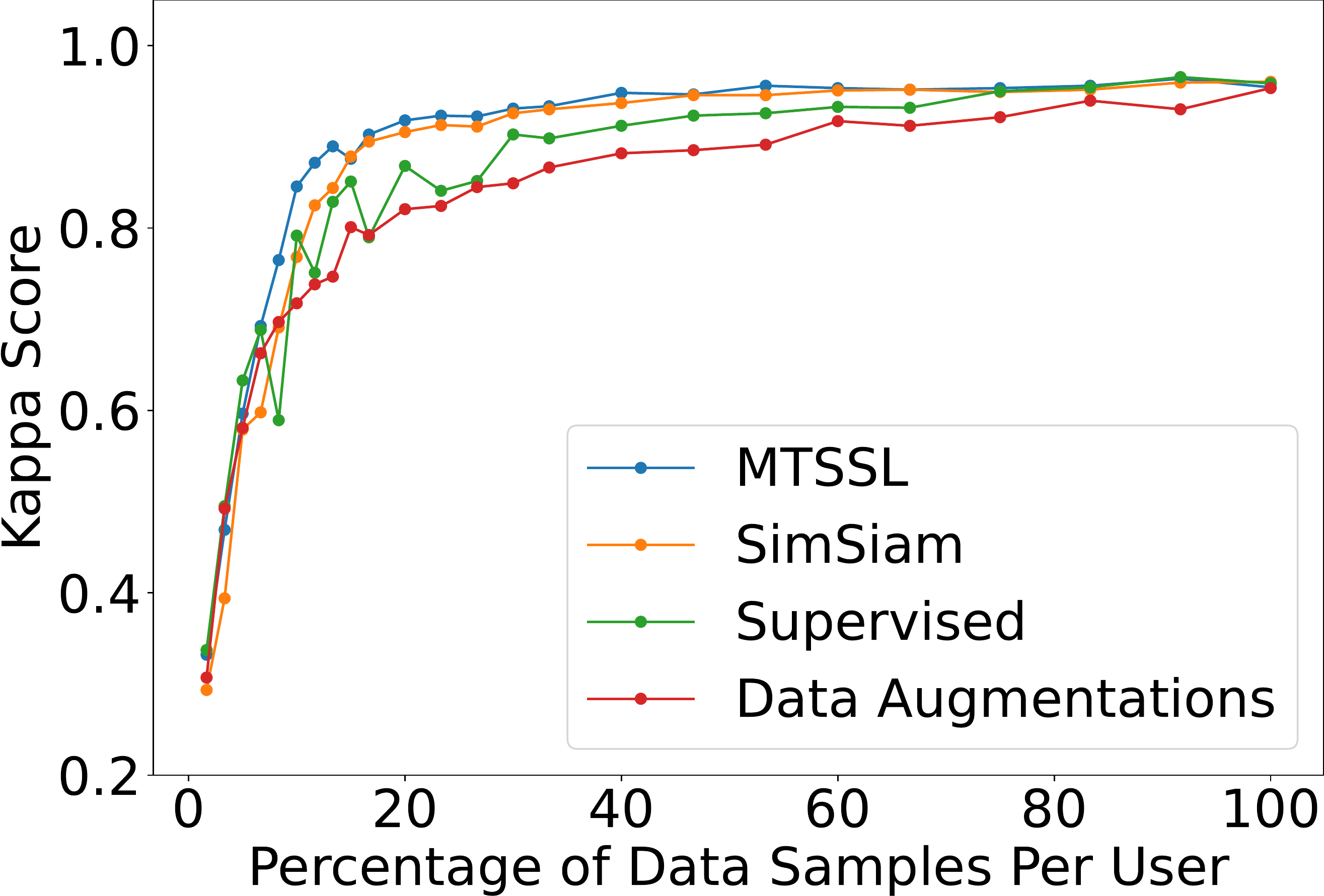}}
\subfloat[MMI]{\label{fig:MMI_3}\includegraphics[trim=0cm 0cm 0cm 0cm, clip=true,scale=0.16]{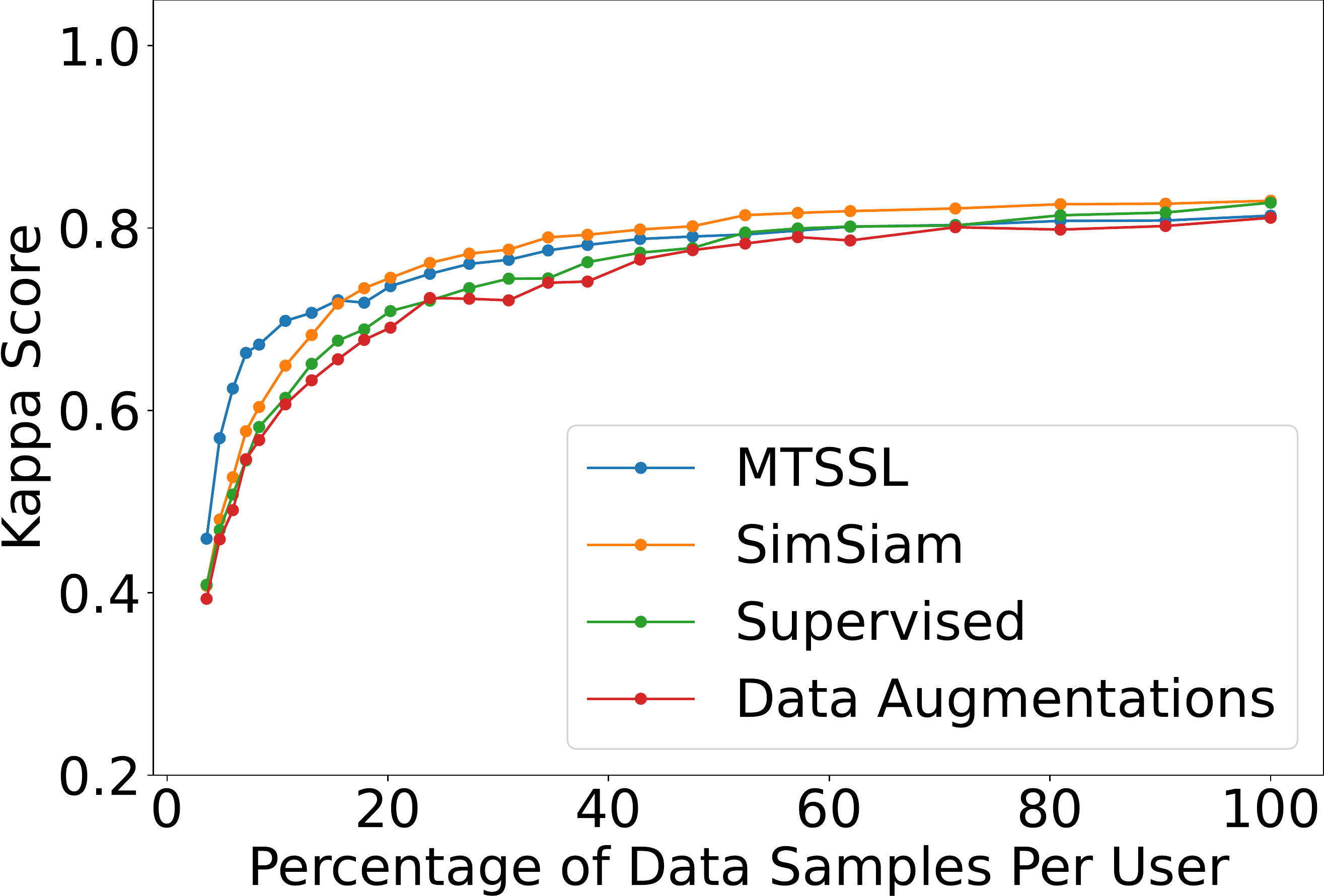}}
\caption{Performance results for Scenario 3}\vspace{-4mm}
\label{fig:scen_3}
\end{figure}

Figure~\ref{fig:MusicID_3} and Figure~\ref{fig:MMI_3} present the results for Scenario 3. Overall, the results are similar to Scenario 2, with SimSiam and multi-task learning performing above the other baselines at lower amounts of labelled data. However, compared to Scenario 2, the performance gap between different methods is reduced in Scenario 3. The reason behind this is having access to a larger amount of labelled data in this scenario compared to the other two scenarios. Here we train the classifier for the total user set. That is, 20 users for \textbf{MusicID} and 109 for \textbf{MMI}. Even though the number of samples a user provides is similar to other two scenarios, when taken as a whole it creates a large labelled dataset. Supervised learning methods benefit from this data and reach a performance similar to self-supervised methods. Even then, multi-task SSL and SimSiam outperform other methods and on the \textbf{MMI} dataset SimSiam slightly outperforms multi-task SSL after 20\% of samples per user.



%% file: sections/ResultsAnalysis.tex
\section{Performance Analysis}
\label{Sec:Analysis}

We next present the results of several other experiments to further analyse the performances of the non-contrastive SSL approach, SimSiam. Since multi-task learning (MTSSL), the other SSL approach among the baselines, also resulted in higher performance compared to traditional supervised learning and data augmentation baselines, where possible, we analyse both SimSiam and MTSSL together. 

Overall, we analyse how different model parameters and data transformations of SimSiam and MTSSL affect the self-supervised learning process. In each experiment, we only change a single variable keeping everything else fixed, and train multiple self-supervised models for different variable values. We conduct these experiments in two settings.

\begin{itemize}
    \item \textbf{Experimental Setting 1 ($Ex_1$)} - We use {\bf Dataset 1} for both pre-training and classifier training. The goal is to assess the quality of the learnt features when pre-training is done with unlabelled data from the same set of users.
    \item \textbf{Experimental Setting 2 ($Ex_2$)} - We do pre-training with \textbf{Dataset 1} and train a classifier on \textbf{Dataset 2} with the objective of assessing the user invariance of the learnt features. 
\end{itemize}

Note that in these two settings, in contrast to Section~\ref{Sec:Results}, we do not fine-tune the feature extractor when training the classifier. This is because we aim to assess the quality of pure features learnt in the self-supervised learning phase. When we fine-tune a model with labelled data, features learnt from an earlier phase can get modified or even overwritten.


We train with a fixed \textit{number of samples per user} - 60 and 300 for the \textbf{MusicID} and \textbf{MMI} datasets, respectively. These two values are chosen according to the data availability of the two datasets. According to Table \ref{tab:datasets}, 68 is the minimum number of samples a user has in the \textbf{MusicID} dataset. Thus, to have a balanced dataset, we use 60 samples per user. Choice of using the maximum possible (and balanced) amount of labelled data is important when evaluating the quality of learnt features to obtain a more generalised view. Similar to \textbf{MusicID} we try to use the maximum balanced dataset for the \textbf{MMI} dataset which turns out to extremely large. At the same time, when we observe Figures \ref{fig:MMI_1}, \ref{fig:MMI_2}, and \ref{fig:MMI_3}, we can observe that after \textit{70\% samples per user}, all the learning methods converge to a single value. That is approximately 300 \textit{percent samples per user} in absolute value for the \textbf{MMI} dataset. Finally, we note that each result we report is the average over 10 experiments to eliminate any biases caused by the random weight initialisation.

\subsection{Transformation Methods}
\label{Sec:Transformations}

\begin{figure*}[t]
\centering \vspace{-3mm}
\subfloat[MusicID]{\label{fig:MusicID_aug}\includegraphics[viewport=65 485 530 725, clip=true,width=0.5\textwidth]{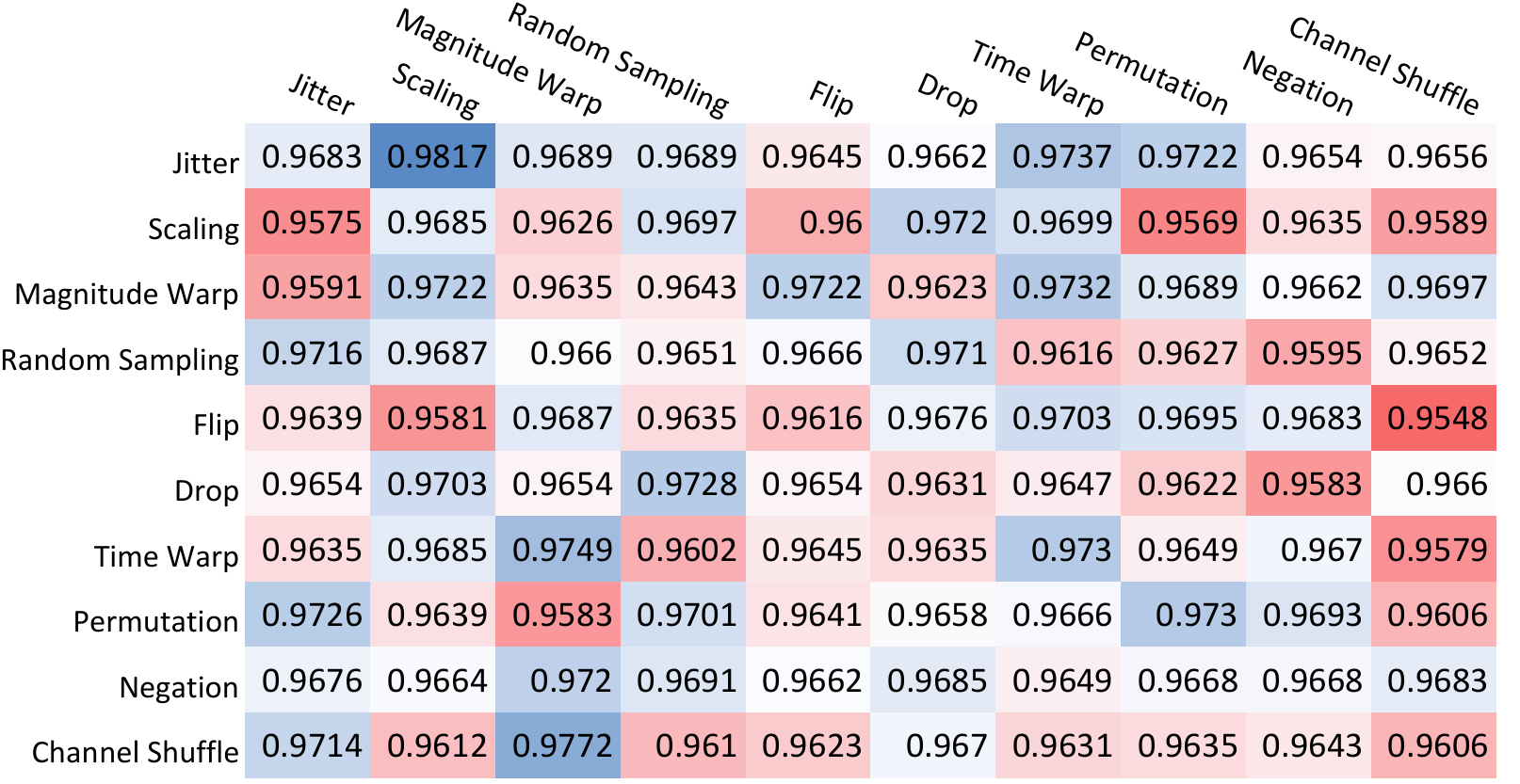}}
\subfloat[MMI]{\label{fig:MMI_aug}\includegraphics[viewport=65 485 530 725, clip=true,width=0.5\textwidth]{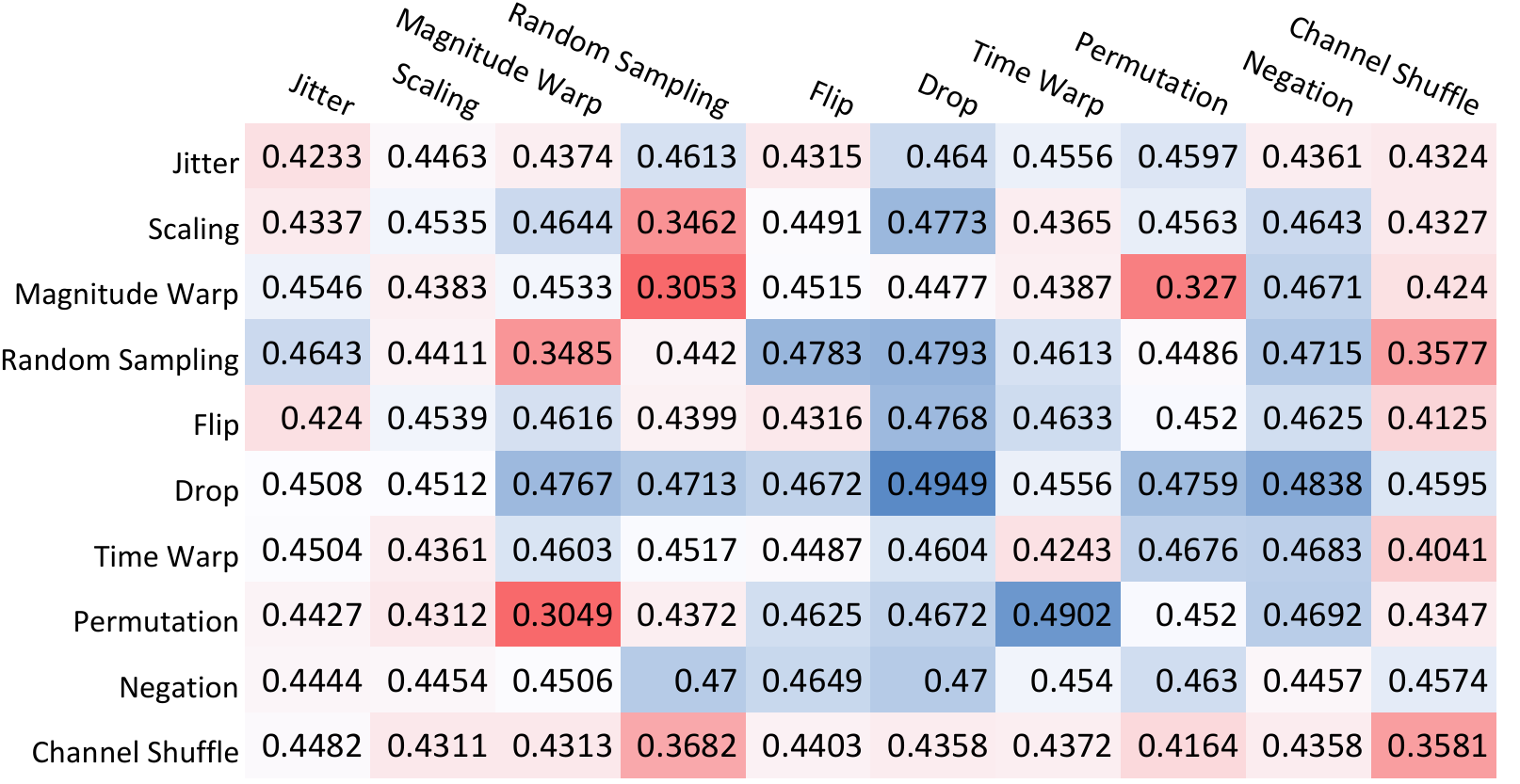}}
\caption{Augmentation comparison for SimSiam method (Kappa scores averaged across  $Ex_1$ and $Ex_2$, ten runs each)}\vspace{-4mm}
\label{fig:aug_simsiam}
\end{figure*}

As described in Section~\ref{Sec:Methodology}, the SimSiam training process involves feeding the network with positive pairs (i.e., two augmented versions of the same input). As such, it is important to identify data transformation/augmentation methods that result in better self-supervised feature learning. Thus, we train the SimSiam network with different augmentation technique pairs and compare their performance. As mentioned earlier, we keep all other network parameters fixed and only change the augmentation technique pair. We evaluate the feature extractors under both the experiment settings, $Ex_1$ and $Ex_2$, for both the datasets and report the average kappa scores in Figure~\ref{fig:MusicID_aug} and Figure~\ref{fig:MMI_aug}.


According to Figure~\ref{fig:MusicID_aug} any data augmentation technique pair gives high kappa scores of over 0.95 for the \textbf{MusicID} dataset. This can be attributed to the smaller size of the \textbf{MusicID} dataset and to the matching experiment conditions of the data collection process, which was designed for authentication applications from the beginning. In contrast, as can be seen from Figure~\ref{fig:MMI_aug} two augmentation technique pairs do not result in high kappa scores for the \textbf{MMI} dataset. Also, there is a considerable variation in kappa scores between pairs. For example, the augmentation pair \textit{Permutation} and \textit{Magnitude Warping} results in only a kappa score of 0.3049 while the pair \textit{Drop} and \textit{Time Warping} results in 0.4902. This can be attributed to the \textbf{MMI} dataset being large and noisy. \textit{Nonetheless, this analysis justifies using more than two augmentation methods to train a better feature extractor in Section~\ref{Sec:Results}. For example, when more augmentations are considered, the kappa scores for the \textbf{MMI} reaches close to 0.8 ({\bf cf.} Section~\ref{Sec:Results})}.




\subsection{Effect of Model Modifications}
\label{SubSec:Modifications}

As mentioned in Section~\ref{Sec:noncontrastive} we did two modifications to the SimSiam learning process to make it more suitable for behavioural biometrics data and improve its performance. Here, we experimentally show how the two modifications we introduced; shallow feature extractor and weight decay, improve the self-supervised training process.

\subsubsection{Depth of the feature extractor}
\label{SubSubSec:depth_fet}

We show that shallow feature extractor networks can lead to higher performance in SimSiam models. We show different feature extractor network configurations we tested in Table~\ref{tab:config}. The first number in each configuration corresponds to the number of filters used in the first 1D convolutional layer and the remaining numbers correspond to the number of filters in concatenated \textit{1D ResNet blocks}. For instance, Configuration 4 of the \textbf{MusicID} dataset corresponds to the architecture illustrated in Figure~\ref{fig:feature} with k=(128,256). We evaluate each of these models using both the experimental setups and report the results in Table~\ref{tab:depth_fet}. The highest accuracies are marked in bold. For comparison, we also report the results of MTSSL when the same feature extractor models are used.

According to Table~\ref{tab:depth_fet}, for SimSiam, Configuration 3 constantly gives the best results. Note that, Configuration 3 layer-wise has the same depth as Configuration 1 and 2. However, Configuration 1 and 2 have a small number of filters compared to Configuration 3. By observing the results, we can conclude that given sufficient convolution filters are there, shallow networks work best with SimSiam. In contrast, the best architecture changes for MTSSL across experimental settings as well as datasets.






\begin{table}[h]
\centering
\smaller
\begin{tabular}{*3l}
\toprule
Config(k) & \textbf{MusicID} & \textbf{MMI}\\\midrule
1 &  32 & 16\\ 
\hline
2 &  64 & 32\\ 
\hline
3 &  128 & 48\\
\hline
4 &  128, 256 & 48, 96\\
\hline
5 &  128, 256, 512 & 48, 96, 192\\
\hline
6 &  128, 256, 512, 1024 & 48, 96, 192, 384\\
\hline
7 &  128, 256, 512, 1024, 2048 & 48, 96, 192, 384, 768\\

\bottomrule
\end{tabular}
\caption{Model configurations for the two datasets}
\label{tab:config}
\end{table}

\begin{table}[h]
\centering
\footnotesize
\begin{tabular}{p{0.2cm}p{0.6cm}p{0.6cm}p{0.6cm}p{0.6cm}p{0.6cm}p{0.6cm}p{0.6cm}p{0.6cm}}
\toprule
 & \multicolumn{4}{c}{\textbf{MusicID}} & \multicolumn{4}{c}{\textbf{MMI}} \\\hline
 & \multicolumn{2}{c}{\textbf{SimSiam}} & \multicolumn{2}{c}{\textbf{MTSSL}} & \multicolumn{2}{c}{\textbf{SimSiam}} & \multicolumn{2}{c}{\textbf{MTSSL}}\\\hline
 & \textbf{$Ex_1$} & \textbf{$Ex_2$}& \textbf{$Ex_1$} & \textbf{$Ex_2$}& \textbf{$Ex_1$} & \textbf{$Ex_2$}& \textbf{$Ex_1$} & \textbf{$Ex_2$}\\\midrule

1 & 0.9461 & 0.9601 & \textbf{0.9709} & 0.9318 & 0.4110 & 0.3796 & 0.5127 & 0.4380\\
\hline 
2 & 0.9751 & 0.9767 & 0.9544 & 0.9534 & 0.4931 & 0.4423 & 0.5827 & 0.5082\\ 
\hline
3 & \textbf{0.9834} & \textbf{0.9966} & 0.9627 & \textbf{0.9750} & \textbf{0.5241} & \textbf{0.4670} & 0.5863 & 0.5160\\
\hline
4 &  0.9482 & 0.9651 & 0.9668 & 0.9534 & 0.4928 & 0.4163 & 0.6238 & 0.5526  \\
\hline
5 &  0.9772 & 0.9518 & 0.9399 & 0.9468 & 0.4396 & 0.3583 & \textbf{0.6532} & 0.5679\\
\hline
6 &  0.9668 & 0.9136 & 0.9419 & 0.9335 & 0.0758 & 0.0430 & 0.6470 & \textbf{0.5733}\\
\hline
7 &  0.8860 & 0.7262 & 0.9192 & 0.9003 & 2e-8 & -2e-8 & 0.6392 & 0.5691\\

\bottomrule
\end{tabular}
\caption{Effect of feature extractor depth}
\label{tab:depth_fet}
\end{table}

\subsubsection{Weight decay}
\label{SubSubSec:weight_dec}
Next, we investigate how high weight decay (a.k.a, regularisation) can improve the performance of self-supervised models. We apply $l_2\;norm$ weight regularisation to the feature extractor network and conduct experiments by varying the regularisation coefficient ($\lambda$) while keeping all other parameters fixed. Table~\ref{tab:reg} shows the results and it is clear that both self-supervised methods benefit from higher weight regularisation.



\begin{table}[h]
\centering
\footnotesize
\begin{tabular}{p{0.6cm}p{0.5cm}p{0.5cm}p{0.5cm}p{0.5cm}p{0.5cm}p{0.5cm}p{0.5cm}p{0.5cm}}
\toprule
 & \multicolumn{4}{c}{\textbf{MusicID}} & \multicolumn{4}{c}{\textbf{MMI}} \\\hline
 & \multicolumn{2}{c}{\textbf{SimSiam}} & \multicolumn{2}{c}{\textbf{MTSSL}} & \multicolumn{2}{c}{\textbf{SimSiam}} & \multicolumn{2}{c}{\textbf{MTSSL}}\\\hline
$\lambda$ & \textbf{$Ex_1$} & \textbf{$Ex_2$}& \textbf{$Ex_1$} & \textbf{$Ex_2$}& \textbf{$Ex_1$} & \textbf{$Ex_2$}& \textbf{$Ex_1$} & \textbf{$Ex_2$}\\\midrule

$0.1$ &  0.9523 & \textbf{0.9036} & 0.9088 & \textbf{0.9236} & 0.5044 & 0.3997 & \textbf{0.6677} & \textbf{0.5859}\\
\hline
$0.01$ &  \textbf{0.9730} & 0.8688 & \textbf{0.9647} & 0.9169 & \textbf{0.5139} & \textbf{0.4148} & 0.6293 & 0.5591\\
\hline
$0.001$ &  0.9523 & 0.8870 & 0.8943 & 0.8571 & 0.4974 & 0.4001 & 0.6080 & 0.5337\\
\hline
$0.0001$ &  0.9255 & 0.8471 & 0.9420 & 0.8503 & 0.5055 & 0.4033 & 0.5885 & 0.5256\\
\hline
$0.00001$ &  0.9357 & 0.8738 & 0.9378 & 0.9019 & 0.5051 & 0.4100 & 0.6077 & 0.5176\\

\bottomrule
\end{tabular}
\caption{Effect of weight decay}
\label{tab:reg}
\end{table}



\subsection{Depth of the Predictor}
\label{SubSec:Predictor}

In contrast to the feature extractor which needs to be shallow for higher performance, our experiments found that the predictor network needs to be deeper for better performance. To show this effect, we conduct several experiments keeping all other parameters fixed and only changing the predictor depth. We present our results in Table~\ref{tab:depth_pred}. Predictor architectures are given in a comma delimited format - corresponds to the dimensions of the Dense layers, from left to right. For example, the sixth predictor architecture corresponds to the architecture illustrated in Figure~\ref{fig:proj_pred}.



\begin{table}[h]
\centering
\footnotesize
\begin{tabular}{l*4c}
\toprule
 & \multicolumn{2}{c}{\textbf{MusicID}} & \multicolumn{2}{c}{\textbf{MMI}} \\\hline
\textbf{Predictor} & \textbf{$Ex_1$} & \textbf{$Ex_2$}& \textbf{$Ex_1$} & \textbf{$Ex_2$}\\\midrule
1. $512$ &  0.9212 & 0.8837 & 0.2424 & 0.1720\\
\hline
2. $2048, 512$ & 0.9440 & 0.9119 & 0.4828 & 0.4041 \\
\hline
3. $4096, 2048, 512$ &  0.9337 & 0.9368 & 0.5008 & 0.4159 \\
\hline
4. $8196, 4096, 2048, 512$ &  0.9399 & 0.9318 & 0.4907 & 0.3993 \\
\hline
5. $8196, 8196, 4096, 2048, 512$ &  0.9337 & \textbf{0.9418} & 0.4860 & 0.3824 \\
\hline
6. $8196, 8196, 8196, 4096, 2048, 512$ &  \textbf{0.9544} & 0.9402 & \textbf{0.5177} & \textbf{0.4263} \\

\bottomrule
\end{tabular}
\caption{Effect of predictor network depth}
\label{tab:depth_pred}
\end{table}


As mentioned in Section~\ref{Sec:noncontrastive}, the role of the predictor is to average the representation vector across all possible augmentations the network has seen. A deeper predictor can memorise more augmentations, consequently making the averaging more precise. During the training process the model compares the output of the predictor $p$ with the output of the encoder network $z$. where $p$ is the mean vector of several augmentations of the same sample and $z$ is the representation vector of a single augmented version of that sample. The task of the network is to make $p$ and $z$ similar. Since $p$ is an averaged vector, in order to make $z$ similar, the encoder network is forced to learn a representation that is common to all the averaged versions. If the predictor is shallow, it can only compare only a few versions of the sample. Therefore, deeper predictor networks can help to learn a more generalised representation.


%% file: sections/Conclusion.tex
\section{Discussion \& Concluding Remarks}
\label{Sec:Conclusion}

Using two EEG-based behavioural biometric datasets and three authentication scenarios, we demonstrated that non-contrastive SSL allows developing label-efficient user classifiers. The SimSiam SSL approach we proposed achieved 4\%--11\% higher on average performance compared to conventional supervised learning and data augmentation baselines. Our approach also resulted in comparable performances to state-of-the-art multi-task SSL approaches in all three scenarios. Next, we discuss the implications of our results, limitations, and possible future extensions. \\ \vspace{-2mm} 

\noindent{{\bf SimSiam and multi-task SSL}} - Though the majority of the time, SimSiam and multi-task learning showed comparable performances, on some occasions, multi-task learning performed better. However, when it comes to training time resource requirements, SimSiam has a distinct advantage in terms of memory footprint over multi-task learning. Multi-task learning requires adding new heads to the network architecture when more transformations are added to the training process. As a result, the network size increases approximately linearly with the number of transformations. In more complex datasets, multi-task learning will require multi-GPU, multi-server distributed training. In contrast, increasing the number of transformations has no impact on the memory footprint of the SimSiam model. The most likely thing that can happen for SimSim is that the number of epochs it needs to be trained may increase with the number of transformations.   \\ \vspace{-2mm} 

\noindent{{\bf Larger dataset and accounting for contextual changes}} - The datasets we explored are relatively homogeneous and stable. That is, the data was collected in similar conditions across sessions. However, contextual changes and biases are major factors in real-world behavioural biometrics systems. Such factors can include user demographics, users' physical condition and activity levels, and the heterogeneity of the hardware used to collect data. In addition to user invariant feature learning, context invariant feature learning will also be required to account for such contextual factors. That means more augmentation techniques need to be investigated, especially those with the potential of being context-invariant, such as augmentation techniques from the frequency domain. 

The true capability of non-contrastive SSL will be more visible when available unlabelled data and the number of users targeted by the authentication application is high. However, the currently publicly available behavioural biometrics datasets for different modalities only have the number of users in the range of a few tens, which is a limitation for further extensions in the area. Also, future work can also explore the potential of non-contrastive SSL on other behavioural biometrics modalities such as gait, typing patterns, and breathing acoustics. \\ \vspace{-2mm} 

\noindent{{\bf Improvements to SimSiam}} - Despite SimSiam-based ideas are relatively new, several subsequent modifications have been proposed to further improve its performance. For example, even though SimSiam architecture avoids network collapsing by using the \textit{stopgradient operation}, it was recently discovered that another phenomenon called \textit{dimensional collapse} can also impact the learning capability of both contrastive learning and non-contrastive learning~\cite{hua2021feature,jing2021understanding}. It is important to analyse such modifications in the context of behavioural biometrics data in particular, and sensor data streams in general to identify possible further improvements. 